\documentclass[12pts,twocolumn,epsf,aps,prb]{revtex4}
\usepackage{float}
\usepackage{color,graphicx}
\usepackage{hyperref}
\begin{document}

\title{Magnetic field induced drastic violation of Wiedemann-Franz law in Dirac semimetal Cd$_{3}$As$_{2}$}

\author{A. Pariari, N. Khan, and P. Mandal}

\affiliation{Saha Institute of Nuclear Physics, 1/AF Bidhannagar, Calcutta 700 064, India}
\date{\today}
\maketitle
\textbf{The journey through the nontrivial band topology beyond the conventional band structure has resulted in the recent discovery of three-dimensional Dirac semimetal phase in Na$_{3}$Bi and Cd$_{3}$As$_{2}$ [1-7]. The bulk state of which is semi-metallic obeying linear energy dispersion, while the surface state is topology protected Fermi arc [1, 2, 8, 9].  Due to the unique band topology, they show different exotic electronic properties of both fundamental and technological interest. From electrical and thermal transport measurements, we have demonstrated a remarkable violation of Wiedemann-Franz law (WFL) under application of magmatic field in Cd$_{3}$As$_{2}$ and the violation becomes more and more drastic with increasing magnetic field strength. Whereas the validity of WFL is the key feature of Landau Fermi-liquid theory in metal, the notion of quasiparticles is the building block to this theory. This implies that the fundamental concept of Landau quasiparticle no longer holds in Cd$_{3}$As$_{2}$ in presence of magnetic field. The continuous break down of Landau quasiparticle framework with field introduces a concept of field induced quantum critical point (QCP) as in the case of heavy fermion compounds YbRh$_{2}$Si$_{2}$ [10], Sr$_{3}$Ru$_{2}$O$_{7}$ [11], etc.}\\

Dirac fermionic excitation and topology introduce different exotic  electronic properties of solids which are now the subject of considerable research interest in condensed matter physics, both from basic research and application point of view. This unique excitation was first discovered in two-dimensional graphene [12] followed by topological insulator [13, 14] and the parent compounds of iron-based superconductors [15]. Unlike conventional fermionic excitations in metals, the quasiparticle excitation in these systems follows a linear energy dispersion. Recently, the three-dimensional (3D) Dirac semimetal state has been theoretically predicted, in which the semimetallic bulk is the 3D analog of graphene  [1, 2, 8, 9]. The 3D Dirac points, where the two Weyl points overlap in momentum space, are protected by the crystal symmetry [1, 2, 8, 9]. Breaking of time reversal symmetry by external magnetic field ($B$) rearranges the Fermi surface [1, 2].\\

Following theoretical prediction [1, 2], 3D Dirac semimetal phase has been discovered experimentally in Na$_3$Bi and Cd$_{3}$As$_{2}$ [3-7]. So far, the major attentions have been focused on the electronic  properties of  Cd$_{3}$As$_{2}$  for understanding the nature of band structure  [16-19]. It has been observed that the Fermi surface of Cd$_{3}$As$_{2}$  undergoes a change with the application of magnetic field [18].  In case of heavy fermion compounds like YbRh$_{2}$Si$_{2}$, the Fermi surface is strongly affected by $B$ [10, 20]. Above a critical field ($B$$_{c}$), the paramagnetic phase is a heavy Fermi liquid with large Fermi surface due to the Kondo effect and the antiferromagnetic phase below $B$$_{c}$ is also a mass enhanced Fermi liquid with small Fermi surface [20]. At $B$=$B$$_{c}$, however, a non-Fermi-liquid ground state emerges in the field-induced QCP  as $T$ approaches towards zero [10]. This motivated us to investigate whether the concept of Landau Fermi-liquid holds across the Fermi surface reconstruction in presence of magnetic field in Cd$_{3}$As$_{2}$, where instead of strong electron-electron correlation, the spin-orbit coupling plays a crucial role [1].\\

Here, we focus on the electrical  and thermal transport properties of Cd$_{3}$As$_{2}$ single crystal in magnetic field. Figure 1 displays the normalized magnetoresistivity ($\rho$) at different temperatures. $\rho$ increases almost linearly with increasing field. Even at 300 K and 9 T magnetic field, $\rho$ shows no sign of saturation, and the resistivity ratio $\rho$$(B)$/$\rho$$(0)$  is$\sim$3. With decreasing temperature, the value of $\rho$$(B)$/$\rho$$(0)$ increases rapidly and at 10 K, it is $\sim$17.5 at 9 T. Inset of Fig. 1 shows temperature dependence of $\rho$. $\rho$ does not follow $T^2$ and linear  $T$ dependence at low temperature and high temperature respectively as in the case of normal metal. \\

\begin{figure}
\includegraphics[width=0.45\textwidth]{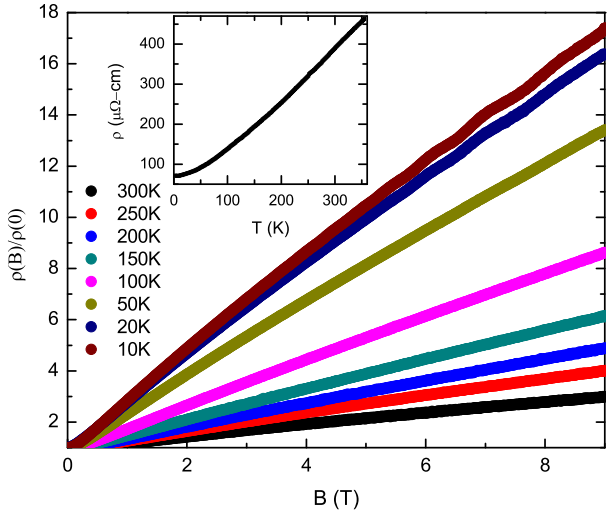}
\caption{(color online) Resistivity ($\rho$) of Cd$_{3}$As$_{2}$ single crystal in presence of magnetic field at some selected temperatures. Inset shows the temperature dependence of $\rho$ in absence of field.}\label{Fig.1}
\end{figure}

In Fig. 2 (a), $\kappa$ has been plotted as a function of $B$ at some selected temperatures between 20 and 300 K.  $\kappa$ decreases very rapidly with increasing $B$ and  tends to saturate at high $B$ which is expected to be the lattice part of thermal conductivity ($\kappa _L$). This saturating behaviour  is more prominent  below 50 K. The inset of Fig. 2 (a) shows temperature dependence of $\kappa$. The  sharp increase in $\kappa$ below 50 K also suggests the smaller proportion of electronic thermal conductivity ($\kappa_e$) to the total at low temperature.  We have fitted the field dependence of thermal conductivity with
\begin{equation}
\kappa = \kappa_L + \frac{\kappa_e}{1+\mu_T^2 B^2}
\end{equation}
to separate $\kappa_e$ from total thermal conductivity $\kappa$, where the $B$$\rightarrow$$\infty$ limit will give the value of $\kappa_L$ and $\mu_T$ is the thermal mobility [21-24]. At low temperature, the extracted value of $ \kappa _L$ is close to the value of $\kappa$ at 9 T. As an example,  at 20 K, we have deduced  $ \kappa _L$$=$3.549 W K$^{-1}$ m$^{-1}$  which is close to the observed value of $\kappa$ at 9 T (3.55 W K$^{-1}$ m$^{-1}$). If one considers the 1st order expansion of the above equation for $\mu$$_{T}$$B$$\ll$ 1 (for very small $B$), one finds $\kappa$-$\kappa_L$=$\kappa_e$($B$)=$\kappa_e$(1-$\mu_T$$^2$$B$$^2$). For a 3D Dirac semimetal, the quadratic $B$ of $\kappa_e$ has been theoretically predicted by considering an energy independent scattering time ($\tau$), which is given by  $\kappa_e$($B$) $\propto [1- (\omega _c \tau)^2]$,  where $\omega _c$ (=$eB$/$\hbar$) is the cyclotron frequency of the electron orbit [25]. This implies that the above assumption of energy independence of scattering time is not valid at high field.\\

\begin{figure}
\includegraphics[width=0.45\textwidth]{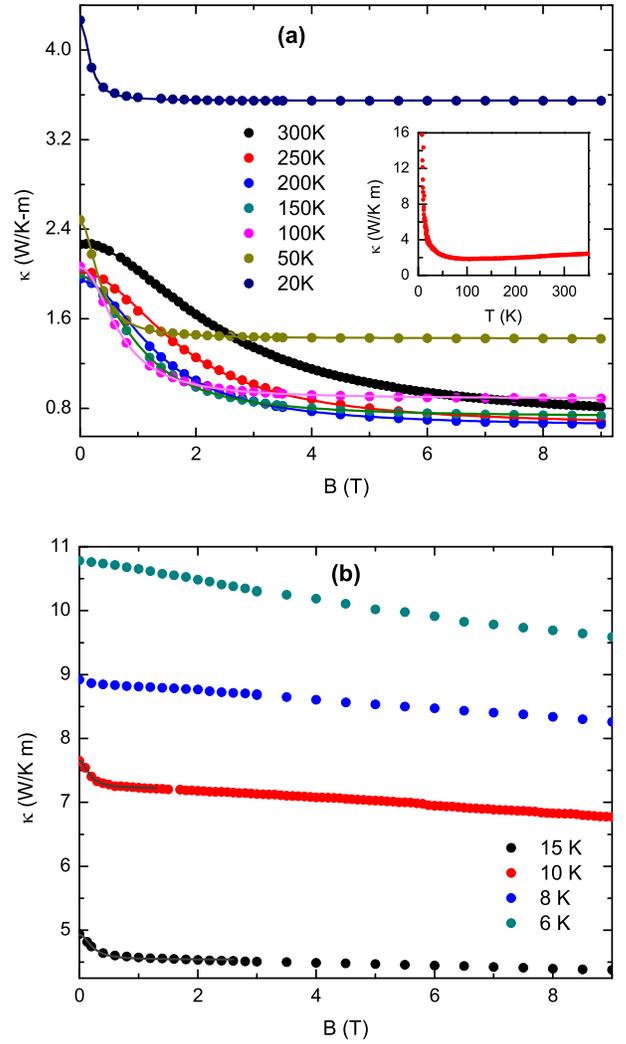}
\caption{(color online) Magnetic field dependence of thermal conductivity (a) From 20 to 300K at some representative temperatures up to 9 T. Solid lines are the fitted to the experimental data. Inset shows the temperature dependence of thermal conductivity at zero field. (b) Below 20 K up to 9 T. It has been fitted by equation (1) at 15 and 10 K over a smaller region (solid lines) relative to higher temperatures data.}\label{Fig.2}
\end{figure}
Below 20 K, instead of showing saturation, $\kappa$ decreases almost linearly up to 9 T and this effect enhances  in the phonon dominated region of $\kappa$, as shown in Fig. 2 (b). Now, equation (1) can not be used to separate $\kappa_L$ from $\kappa$. We have measured  magnetostriction at low temperature to check whether $B$ induces any structural change. But we have not observed any noticeable changes  in lattice parameters. The specific heat does not show any $B$ dependence which implies that the phonon density of state is not affected by $B$. So, we propose that the observed phenomenon is due to the reduction in lattice thermal conductivity in magnetic field.  Recently, the lattice thermal conductivity of diamagnetic semiconductor InSb is  reported to exhibit $B$ dependence [26]. In that work, the authors have argued that $B$ creates a spatial gradient in diamagnetic moment around the displaced atoms in phonon vibration which exerts an anharmonic magnetic force on the displaced atom and as a result, a reduction in lattice thermal conductivity.  We have estimated about 10\% reduction in $\kappa_L$ at 6 K and 9 T which is comparable to that for InSb [26].\\

To calculate $\kappa$$_{e}$/$\sigma$$T$, i.e, the Lorentz number ($L$), we have extracted $\kappa_e$ from $\kappa$ using equation 1. Figure 3 (a) displays the temperature variation of $L$ in absence of magnetic field.  $L$ over most of the temperature range is only about 70\% of the ideal value L$ _0$ (2.44$\times$10$^{-8}$ $W$$\Omega$K$^{-2}$). $L$ shows a shallow dip around 100 K and  approaches rapidly to L$_0$  at low temperature. Below 10 K, it is difficult to calculate the precise value of $L$ because $\kappa_L$ is affected by $B$. The temperature dependence of $L$ is qualitatively similar  to earlier report [22], where the $B$ dependence of $\kappa$ was measured only up to 2 T to separate $\kappa_e$ using equation 1 and to calculate $L$.\\

\begin{figure}
\includegraphics[width=0.45\textwidth]{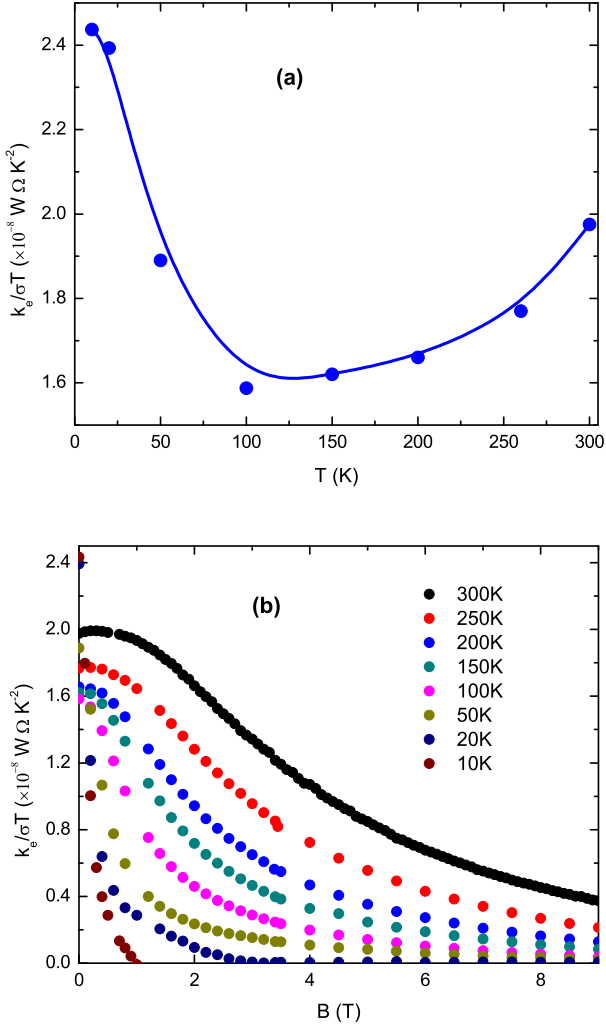}
\caption{(color online) (a) Temperature dependence of Lorentz number (L), $\kappa$$_{e}$/$\sigma$$T$, at zero field. (b) Reduction of $\kappa$$_{e}$/$\sigma$$T$ ratio under application of magnetic field upto 9 T at some constant temperatures.}\label{Fig.3}
\end{figure}
Figure 3 (b) shows $\kappa$$_{e}$/$\sigma$$T$ ratio under application of $B$ up to 9 T at some selected temperatures in the range 10 to 300 K. $\kappa$$_{e}$/$\sigma$$T$ decreases very rapidly with increasing $B$.  At 300 K, it decreases from 1.95 to 0.37 as  $B$ increases from zero to 9 T. At low temperatures, however, $L$ reduces more rapidly with increasing field strength. For example, at 20 K, $\kappa$$_{e}$/$\sigma$$T$ drops from its zero field value 2.39 to $\sim$0.006, only at 3 T. There may be some error in the calculated value of $L$ at 10 K due to smaller proportion of $\kappa$$_{e}$ and the effect of $B$ on $\kappa_L$ but the qualitative behavior of $L(B)$ remains same. At high fields and low temperature, $L$ is found to be few orders of magnitude smaller than L$_0$. This clearly demonstrates a drastic violation of the Wiedemann-Franz law  under application of magnetic field in Dirac semimetal Cd$_{3}$As$_{2}$. The violation of this universal law can be observed from the non-identical $B$ dependence of $\kappa$ and $\sigma$.  $\rho$$(B)$ does not follow the predicted quadratic $B$ dependence,    $\rho(B)/\rho(0)={1+\mu_T^2 B^2}$, but linear in $B$ [23, 24]. As a result of such unusual linear $B$ dependence of $\rho$, the ratio $\kappa_e$($B$)/$\sigma(B)$  fails to remain constant. \\

In order to understand the effect of $B$ on the Fermi surface, we have measured the Seebeck coefficient ($S$). $S$ increases monotonically with $B$ and tends to saturate at high $B$ as shown in Fig. 4. Also with decreasing temperature, the value of $S$ decreases monotonically and tends to saturate at a relatively lower field strength. The well known Mott's semiclassical formula of thermoelectric power, $S=\frac{\pi^2}{3e}\frac{T}{\sigma(\mu)}\frac{\partial\sigma(\varepsilon)}{\partial\varepsilon}|_{\varepsilon=\mu}$, has been used in graphene to probe the relaxation process [27]. Here $\sigma(\varepsilon)$ is energy dependent conductivity, $\mu$ is the chemical potential and $\mu$$=$$E_F$ for $T$$<<$$T_F$. Using the similar algorithm in case of "3D analog of graphene" and scattering time ($\tau$)$\propto \varepsilon^m$, the above equation simplifies to, $S=\frac{\pi^2k_B}{3e}\frac{k_BT}{E_F}(m+2)$. From the linear $T$ dependence of $S$, we have deduced scattering exponent $m$$\sim$0.15 for the relaxation process in Cd$_{3}$As$_{2}$ single crystal with Fermi energy $\sim$270 meV [28]. This linear $T$ dependence of $S$ is robust under application of field (see Supplementary Information). Now, considering the above equation and Fig. 4, it can be shown that $m$ increases monotonically with $B$  and tends to saturate at 1 at high field. Also, it is apparent from Fig. 4 that $m$ reaches close to 1 at lower $B$ with decreasing temperature. The value of $m$ determines the nature of relaxation process of the charge carrier. For examples, $m$=2 for screened charged impurity scattering,  $m$=-2 for short-range disorder [25] and for unscreened charged impurity scattering $m$ is 1 [27], etc. The strong $B$ dependence of $m$ can be ascribed to field induced change of Fermi surface, similar to the earlier report [18].\\
\begin{figure}
\includegraphics[width=0.45\textwidth]{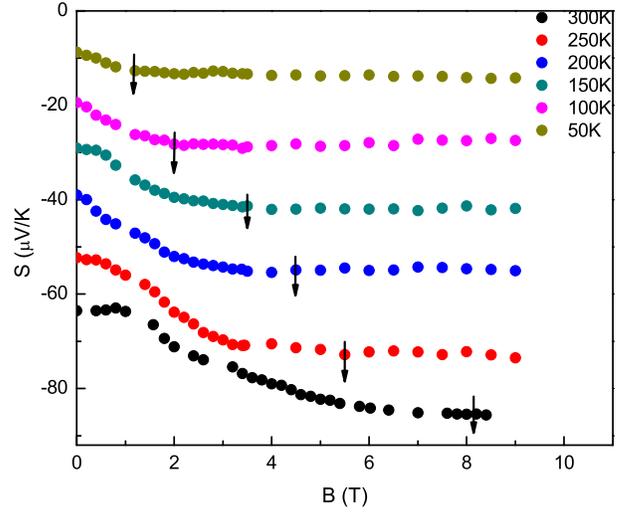}
\caption{(color online) Figure shows the effect of magnetic field on Seebeck coefficient ($S$) upto 9 T at some representative temperatures between 20 to 300 K. Right side of downward arrow represents the value of m close to 1.}\label{Fig.4}
\end{figure}
\begin{figure}
\includegraphics[width=0.45\textwidth]{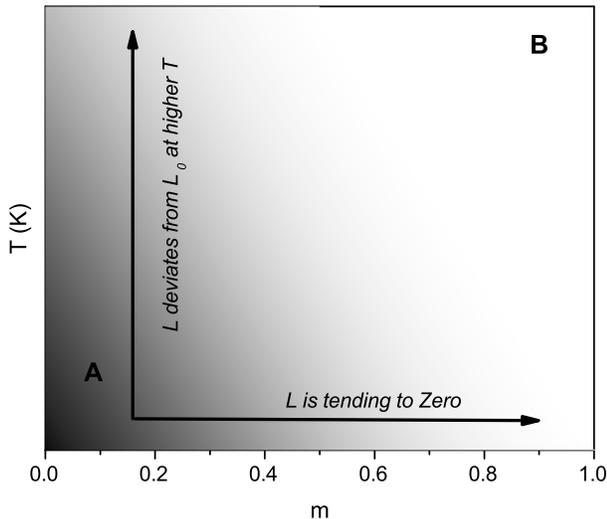}
\caption{(color online) A schematic representation. Here, \textbf{A} representing the region where the value of Lorentz number ($L$) is larger than in region \textbf{B}. Deeper shaded region represents the value of L closer to $L_0$ and at (0,0), L=$L_0$. Also it should be noted that there is no sharp boundary between region \textbf{A} and \textbf{B}. System continuously evolves from region \textbf{A} to region \textbf{B} with the variation of magnetic field and temperature.}\label{Fig.5}
\end{figure}

With increasing magnetic field strength,  the scattering exponents $m$ tends to saturate at 1 while the Lorentz number $L$ falls drastically from its ideal value $L_0$. This implies a continuous violation of Wiedemann-Franz law (WFL), i.e, the Landau quasiparticle no longer holds under application of magnetic field. However, the present and earlier studies [22] clearly show that $L$ approaches  L$ _0$ at low temperature and zero field. Based on our experimental observation, we have constructed a schematic diagram in Fig. 5. \textbf{A} represents the shaded region, where $L$ is closer to $L_0$ than the value in region \textbf{B} and at the origin (0,0), $L$=$L_0$. The system continuously evolves from \textbf{A} to \textbf{B} and vice versa  with the variation of $m$($B$). Now, by tuning $m$ from $\sim$ 0  to 1 through magnetic field, one observes a continuous breakdown of Landau Fermi liquid behaviour.  Thus, $m$=1 looks like a quantum critical point (QCP) at $T$ tends to zero limit. As if, we have been able to achieve the one side of the point. The other side can not be accessed by tuning $B$ as $m$ saturates at 1. Similar kind of behaviour has been observed in heavy fermion compound like YbRh$_{2}$Si$_{2}$ [10], in which $B$ induces a QCP between paramagnetic heavy Fermi-liquid state and antiferromagnetic mass enhanced Fermi-liquid state, where non-Fermi-Liquid ground state emerges. The present study introduces the concept of field-induced QCP in Dirac semimetal Cd$_{3}$As$_{2}$, where spin-orbit coupling is more important than electron-electron interaction. A qualitatively similar kind of behaviour has been predicted   using the renormalization group analysis [29]. In that work, the authors have claimed that a quantum phase transition in 3D Dirac semimetals occurs due to the fascinating interplay of Coulomb interaction and disorder. The non-Fermi liquid  ground state can be achieved by increasing the strength of disorder.  This theoretical study is consistence with our observation. It seems that $B$ is playing the role of disorder in the present study.  As an another possible explanation, the field-induced violation of conventional WFL may be an interesting property of topological Fermi-liquid [30](Landau Fermi-liquid including Berry's phase in presence of broken inversion or time-reversal symmetry), which is unexplored till now. A quadratic $B$ dependence of Lorentz number has been theoretically predicted for Weyl semimetal [31] (which is also a topological Fermi-liquid system), due to the chiral anomaly. Further theoretical studies are necessary to understand the possible origin of such a strong violation of WFL.\\

{\large \textbf{Methods}}\\Single crystals of Cd$_{3}$As$_{2}$ were synthesized by  chemical vapor transport technique [28]. Powder X-ray diffraction of crashed single crystals shows that these crystals have I$_{41}$/acd space group and contain no impurity phases [28]. The Cd$_{3}$As$_{2}$ single crystal was cut and polished to a bar shape, with dimension$\sim$3$\times$2$\times$0.6$\times$ $mm^3$ for transport measurements.  The  resistivity, thermal conductivity and thermoelectric power (Seebeck coefficient) measurements on Cd$_{3}$As$_{2}$ single crystals were done by four-probe technique in a physical property measurement system (Quantum Design) upto 9 T. Though several single crystals have been studied, we present the data for a single crystal as a representative. Qualitative similar behavior has been observed for other crystals. Creating a temperature gradient along the current direction, the measurements were performed by applying magnetic field perpendicular to them.\\

{\large \textbf{References}}\\

1. Wang,  Z. J., Sun, Y., Chen, X. Q., Franchini, C., Xu,  G., Weng, H. M., Dai, X., and Fang, Z.  Dirac semimetal and topological phase transitions in A$_{3}$Bi (A = Na, K, Rb). \emph{Phys Rev B} \textbf{85}, 195320 (2012).

2. Wang, Z., Weng, H., Wu, Q., Dai, X. \& Fang, Z. Three-dimensional Dirac semimetal and quantum transport in Cd$_{3}$As$_{2}$. \emph{Phys. Rev.B} \textbf{88}, 125427 (2013).

3. Liu, Z. K., Zhou, B., Zhang, Y., Wang, Z. J.,  Weng, H. M., Prabhakaran, D., Mo, S.- K., Shen, Z. X., Fang, Z., Dai, X., Hussain, Z. \&  Chen, Y. L. Discovery of a Three-Dimensional Topological Dirac Semimetal, Na$_{3}$Bi.  \emph{Science} \textbf{343}, 864 (2014).

4. Xu, S. -Y., Liu, C., Kushwaha, S. K., Chang, T. -R., Krizan, J. W., Sankar, R., Polley, C. M., Adell, J., Balasubramanian, T., Miyamoto, K., Alidoust, N., Bian, G., Neupane, M., Belopolski, I., Jeng, H. -T.,  Huang, C. -Y., Tsai, W. -F., Lin, H.,  Chou, F. C., Okuda, T., Bansil, A., Cava, R. J. \& Hasan, M. Z. Observation of a bulk 3D Dirac multiplet, Lifshitz transition, and nestled spin states in Na$_{3}$Bi. \emph{Science} \textbf{347}, 294 (2015).

5. Liu, Z. K.,  Jiang, J., Zhou, B., Wang, Z. J., Zhang,Y., Weng, H. M.,  Prabhakaran, D., Mo, S. -K., Peng, H., Dudin, P., Kim, T., Hoesch, M., Fang, Z., Dai, X., Shen, Z. X., Feng, D. L., Hussain, Z. \& Chen, Y. L. A stable three-dimensional topological Dirac semimetal Cd$_{3}$As$_{2}$. \emph{Nat. Mater.} \textbf{13}, 677 (2014).

6. Neupane, M., Xu, S. -Y., Sankar, R., Alidoust, N., Bian, G., Liu, C., Belopolski, I., Chang, T. -R., Jeng, H. -T., Lin, H., Bansil, A., Chou, F.  \& Hasan, M. Z. Observation of a three-dimensional topological Dirac semimetal phase in high-mobility Cd$_{3}$As$_{2}$. \emph{Nat. Commun.} \textbf{5}, 3786 (2014).

7. Borisenko, S., Gibson, Q., Evtushinsky, D., Zabolotnyy, V., B\"{u}chner, B. \& Cava, R. J. Experimental Realization of a Three-Dimensional Dirac Semimetal. \emph{Phys. Rev. Lett.} \textbf{113}, 027603 (2014).

8. Young, S. M., Zaheer, S., Teo, J. C. Y., Kane, C. L., Mele, E. J. \& Rappe, A. M. Dirac Semimetal in Three Dimensions. \emph{Phys. Rev. Lett.} \textbf{108}, 140405 (2012).

9. M\~{a}nes, J. L. Existence of bulk chiral fermions and crystal symmetry. \emph{Phys. Rev. B} \textbf{85}, 155118 (2012).

10. Pfau, H., Hartmann, S., Stockert, U., Sun, p., Lausberg, s., Brando, m., Friedemann, s., Krellner, C., Geibel, C., Wirth, s., Kirchner, s., Abrahams, E., Si, Q. \& Steglich, F.  Thermal and electrical transport across a magnetic quantum critical point. \emph{Nature} \textbf{484}, 493-497 (2012).

11. Grigera, S. A. et al. Magnetic field-tuned quantum criticality in the metallic ruthenate Sr$_{3}$Ru$_{2}$O$_{7}$. \emph{Science} \textbf{294}, 329-332 (2001).

12. Castro Neto, A. H., Peres, N. M. R., Novoselov, K. S. \& Geim, A. K. The electronic properties of graphene. \emph{Rev. Mod. Phys.} \textbf{81}, 109(2009).

13. Hasan, M. Z. \& Kane, C. L. \emph{Colloquium}: Topological insulators. \emph{Rev. Mod. Phys.} \textbf{82}, 3045 (2010).

14. Qi, X. -L. \&  Zhang, S. -C. Topological insulators and superconductors. \emph{Rev. Mod. Phys.} \textbf{83}, 1057 (2011).

15. Richard,P., Nakayama,K. , Sato,T., Neupane,M., Xu,Y.-M., Bowen,J. H., Chen, G. F., Luo, J. L., Wang, N. L., Dai, X., Fang, Z., Ding, H. \& Takahashi, T. Observation of Dirac Cone Electronic Dispersion in BaFe$_{2}$As$_{2}$ \emph{Phys. Rev. Lett.} \textbf{104}, 137001 (2010).

16. He, L. P., Hong, X. C., Dong, J. K.,  Pan, J., Zhang, Z.,  Zhang, J. \&  Li, S. Y. Quantum Transport Evidence for the Three-Dimensional Dirac Semimetal Phase in Cd$_{3}$As$_{2}$. \emph{Phys. Rev. Lett.}  \textbf{113}, 246402 (2014).

17. Potter, A. C., Kimchi, I., \& Vishwanath, A. Quantum oscillations from surface Fermi arcs in Weyl and Dirac semimetals. \emph{Nature Communications} \textbf{5}, 5161 (2014).

18. Liang, T., Gibson, Q.,  Ali, M. N., Liu, M., Cava, R. J. \& Ong, N. P. Ultrahigh mobility and giant magnetoresistance in the Dirac semimetal Cd$_{3}$As$_{2}$. \emph{Nat. Mater.} \textbf{14}, 280 (2015).

19. Feng, J., Pang, Y., Wu, D., Wang, Z., Weng, H., Li, J., Dai, X., Fang, Z., Shi, Y. \& Luyar, L. Large linear magnetoresistance in Dirac semi-metal Cd$_{3}$As$_{2}$ with Fermi surfaces close to the Dirac points, \emph{arXiv}:1405.6611.

20. Gegenwart, P. et al. Magnetic-field induced quantum critical point in YbRh2Si2. \emph{Phys. Rev. Lett.} \textbf{89}, 056402 (2002).

21. Korenblit, L. L. \& Sherstobitov, V. E. Electron scattering insb-type semiconductors. \emph{Soviet physics-semiconductors} \textbf{2}, 564 (1968).

22. Armitage, D. \& Goldsmid, H. J. The thermal conductivity of cadmium arsenide. \emph{J . Phys. C }, \emph{VOL.} 2, 2138 (1969).

23. Lukas, K. C., Liu, W. S., Joshi, G., Zebarjadi, M., Dresselhaus, M. S., Ren, Z. F., Chen, G., \& Opeil, C. P. Experimental determination of the Lorenz number in Cu$_{0.01}$Bi$_{2}$Te$_{2.7}$Se$_{0.3}$ and Bi$_{0.88}$Sb$_{0.12}$. \emph{Phys. Rev. B} \textbf{85}, 205410 (2012).

24. Jacoboni, C. {\it Theory of Electron Transport in Semiconductors} (Springer, Berlin, 2010).

25. Lundgren, R., Laurell, P. \& Fiete, G. A. Thermoelectric properties of Weyl and Dirac semimetals. \emph{Phys. Rev. B} \textbf{90}, 165115 (2014).

26. Jin, H., Restrepo, O. D., Antolin, N., Boona., S. R., Wind, W., Myers, R. C., and Heremans, J. P. Phonon-induced diamagnetic force and its effect on the lattice thermal conductivity. \emph{Nat. Mater.} \textbf{14}, 601-606 (2015).

27. Hwang, E. H., Rossi, E. \& Das Sarma, S. Theory of thermopower in two-dimensional graphene. \emph{Phys. Rev. B}  \textbf{80}, 235415 (2009).

28. Pariari, A., Dutta, P. \& Mandal, P. Probing the Fermi surface of three-dimensional Dirac semimetal Cd$_{3}$As$_{2}$ through the de Haas–van Alphen technique. \emph{Phys. Rev. B} 91, 155139 (2015).

29. Moon, E.- G.  \&  Kim, Y. B. Non-Fermi Liquid in Dirac Semi-metals. \emph{arXiv}:1409.0573.

30. Haldane, F. D. M. Berry Curvature on the Fermi Surface: Anomalous Hall Effect as a Topological Fermi-Liquid Property. \emph{Phys. Rev. Lett.}  \textbf{93}, 206602 (2004).

31. Kim, K-S. Role of axion electrodynamics in a Weyl metal: Violation of Wiedemann-Franz law. \emph{Phys. Rev. B} \textbf{90}, 121108(R) (2014).

\textbf{Acknowledgement:} We thank A. Midya, and A. Paul for their help during measurements and useful discussions.\\

\pagebreak
\large \textbf{Supplementary information for "Magnetic field induced drastic violation of Wiedemann-Franz law in Dirac semimetal Cd$_{3}$As$_{2}$"}
\section{Temperature dependence of Seebeck coefficient ($S$) under application magnetic field ($B$).}
\begin{figure}[h!]
\includegraphics[width=0.55\textwidth]{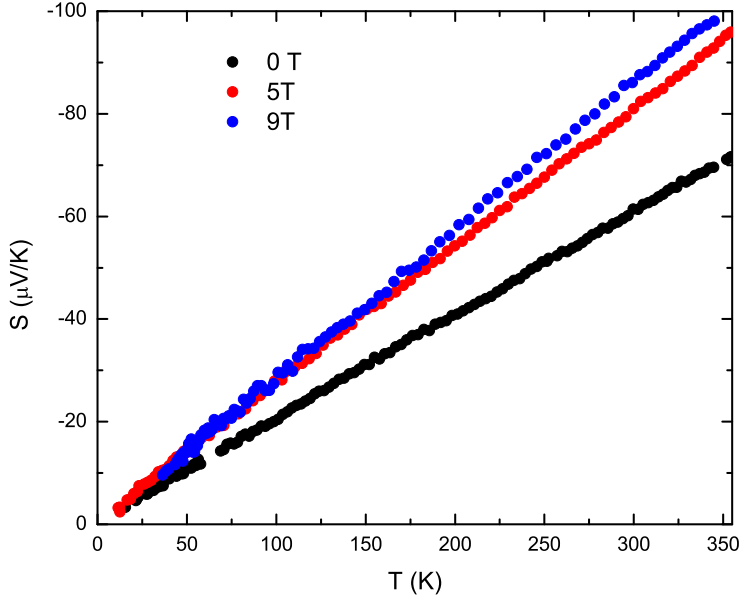}
\caption{(Color online) Temperature dependence of $S$ at 0T, 5T and 9T magnetic field.}\label{rh}
\end{figure}
Figure 6 shows linear T dependence of $S$  both in presence and absence of external magnetic field ($B$) upto 350K. This implies Mott's semiclassical formula of thermoelectric power also holds in presence of $B$. So the expression, $S=\frac{\pi^2k_B}{3e}\frac{k_BT}{E_F}(m+2)$, can be used to calculate the $B$ dependence of scattering exponent ($m$).\\

\end{document}